\documentclass[]{article}
\usepackage{amsmath, mathrsfs, amssymb,amsfonts,amsthm,graphicx, epsf, dcolumn, yfonts, }
\usepackage[hyperfootnotes=true]{hyperref}
\usepackage[normalem]{ulem}
\usepackage{color}
\usepackage{authblk}
\usepackage{slashed}
\usepackage{setspace}
\usepackage[a4paper, total={6in, 8in}]{geometry}
\usepackage{cancel}
\usepackage{wasysym}
\usepackage[lofdepth,lotdepth]{subfig}
\usepackage{float}
\usepackage{dsfont}
\usepackage[utf8]{inputenc}
\usepackage{ragged2e}
\newcommand{\tr}[0]{\textrm{tr}}

\title{Fixing semi-classical physics from first principles:\\ how to derive effective classical-quantum dynamics\\ from open quantum theory \footnote{This is based on a talk given at Concepts of Quantum and Spacetime, KEK, March 2026 and is intended as a lighter companion to \cite{layton2022healthier,layton2024classical}.}}

\date{April 2026}
\author[1]{Isaac Layton}
\affil[1]{Department of Applied Physics, The University of Tokyo,
7-3-1 Hongo, Bunkyo-ku, Tokyo 113-8656, Japan}

\begin{document}

\maketitle

\begin{abstract}

Semi-classical approaches approximate fully quantum descriptions with partially classical ones. Here we use a toy model to highlight the failings of the standard mean-field semi-classical approach, and show how including environmental decoherence can lead to improved semi-classical theories that are exact descriptions of the original quantum dynamics. In doing so, we show how consistent models of classical-quantum dynamics can arise as effective descriptions of open quantum systems.

\end{abstract}


\subsection*{Introduction}

Given the complexity of the quantum world, it is almost always necessary to approximate part of it classically. At its most basic level, this means approximating a bipartite quantum system by an effective picture with both classical and quantum variables i.e.
\begin{equation} \label{eq: QQ->CQ}
    \textrm{quantum} \leftrightarrow \textrm{quantum} \quad  \approx \quad \textrm{classical} \leftrightarrow \textrm{quantum},
\end{equation} where $\leftrightarrow$ denotes an interaction. Such descriptions are known as semi-classical.

The most famous semi-classical theory is arguably mean-field semi-classical gravity \cite{rosenfeld1963quantization,moller1962theories}. Here a full, and as yet unknown, theory of quantum gravity and quantum matter is approximated by the following system of dynamics
\begin{equation} \label{eq: semi_classical_einstein}
    G_{\mu\nu}=\frac{8\pi G}{c^4}\langle \hat{T}_{\mu\nu} \rangle \quad + \quad \quad \textrm{quantum-field theory in curved space}.
\end{equation} However, this semi-classical dynamics is also often criticised, for violating a number of basic physical principles \cite{eppley1977necessity,gisin1990weinberg} (at least in spirit, if not in practice \cite{Kent2005,fedida2026thermodynamics}), and having limited regimes of validity \cite{Hu:2008rga}.

In this work, we aim to restore the reputation of semi-classical theories, by highlighting that the dynamics provided by \eqref{eq: semi_classical_einstein} are not the only way of approximating quantum dynamics. By following first principles and starting from a fully quantum model, we show that one can find semi-classical descriptions that are both consistent and with much broader regimes of validity.

\subsection*{Toy model} \label{sec: linear_oscillators}

To make this concrete, we consider a toy model of two interacting quantum systems, which we call $C$ and $Q$. The $C$ system will be taken to be a quantum particle, with the usual position and momentum operators satisfying $[\hat{q},\hat{p}]=i\hbar$. The $Q$ system will be a qubit, with a Hermitian operator $\hat{A}=\hat{A}^\dag$. We will consider the systems to be linearly coupled, evolving under the unitary dynamics
\begin{equation} \label{eq: toy_model_unitary_dynamics}
    \frac{d\hat{\rho}_{CQ}}{dt}=-\frac{i}{\hbar}[\lambda \hat{q} \hat{A},\hat{\rho}_{CQ}],  
\end{equation} where $\lambda$ is the coupling strength and $\hat{\rho}_{CQ}$ denotes the bipartite density operator. Note that for simplicity of presentation we work in a large mass limit such that we drop the usual kinetic term of the quantum particle.

To study how semi-classical approximations apply to this system, we make use of the partial Wigner representation \cite{boucher1988semiclassical, kapral1999mixed,layton2024classical}. First, note that for a single system, the standard Wigner function $W(q,p)$ is obtained by mapping the Hilbert space $\mathcal{H}_C$ to phase space  $\mathcal{M}$
\begin{equation}
 W(q,p)=\tr(\hat{\rho}_{C}\hat{A}_{q,p}) 
\end{equation} where here $\hat{A}_{q,p}=\frac{1}{\pi\hbar}\int_{-\infty}^\infty  e^{-\frac{i}{\hbar}p y} |q+\frac{y}{2}\rangle \langle q-\frac{y}{2}| dy$. For a bipartite quantum system, the partial Wigner function $\hat{\varrho}^W(q,p)$ is obtained by mapping $\mathcal{H}_C$ into phase space, whilst leaving the $\mathcal{H}_Q$ Hilbert space intact i.e.
\begin{equation}
      \hat{\varrho}^W(q,p)=\tr_C(\hat{\rho}_{CQ}\hat{A}_{q,p}), 
\end{equation} where here only the \textit{partial} trace is performed. Correspondingly, while $W(q,p)$ is a real-valued function of phase space, $\hat{\varrho}^W(q,p)$ is operator valued. Applying this mapping to \eqref{eq: toy_model_unitary_dynamics} we find the dynamics
\begin{equation} \label{eq: toy_model_unitary_dynamics_partial_Wigner}
    \frac{\partial \hat{\varrho}^W}{\partial t}=-\frac{i}{\hbar}[\lambda q \hat{A}, \hat{\varrho}^W] + \frac{\lambda}{2}\{\hat{A},\frac{\partial\hat{\varrho}^W}{\partial p}\}_+,
\end{equation} which provides an entirely equivalent description of the unitary quantum dynamics \eqref{eq: toy_model_unitary_dynamics} in a partial phase space picture. The utility of this representation comes from the fact that while effective classicality of a system implies ${W(q,p)\geq 0}$, effective classicality of a subsystem implies ${\hat{\varrho}^W(q,p)\succeq 0}$ i.e. positive semi-definite as an operator \cite{layton2024classical}.

\subsection*{Standard semi-classical approach}

Equipped with a partial phase space representation \eqref{eq: toy_model_unitary_dynamics_partial_Wigner} of our fully quantum dynamics \eqref{eq: toy_model_unitary_dynamics}, we now turn to see how to arrive at a semi-classical evolution law.

The first method we present is the standard one. Namely, we assume that the quantum states of $C$ and $Q$ approximately factorise at all times  i.e.
\begin{equation} \label{eq: factorisation}
    \hat{\varrho}^W(q,p,t)\approx W(q,p,t) \hat{\rho}(t) \quad \quad \forall t,
\end{equation} where here $\hat{\rho}$ denotes a state on $\mathcal{H}_Q$. Taking the phase and Hilbert space traces of \eqref{eq: toy_model_unitary_dynamics_partial_Wigner}, we see that this implies 
\begin{equation} \label{eq: semi_classical_liouville}
    \partial_t W(q,p,t)=\tr\bigg[\frac{\partial \hat{\varrho}^W}{\partial t}\bigg]\approx\lambda \langle \hat{A} \rangle \frac{\partial W(q,p,t)}{\partial p}
\end{equation}
\begin{equation}
    \partial_t \hat{\rho}(t) = \int dq dp\big[ \partial_t \hat{\varrho}^W(q,p,t) \big] \approx \int dq dp W(q,p,t)\bigg(-\frac{i}{\hbar}[\lambda q \hat{A},\hat{\rho}(t)]\bigg),
\end{equation} where $\langle \hat{A} \rangle =\tr (\hat{A}\hat{\rho})$ \cite{gerasimenko2007comment}. When $W(q,p)\geq0$, these correspond to the classical Liouville equation under a force $-\lambda \langle \hat{A} \rangle$, and unitary quantum dynamics averaged over $W(q,p)$, and hence can be unravelled along individual trajectories as
\begin{equation} \label{eq: toy_model_standard_semi-classical}
    \frac{dp}{dt}=-\lambda \langle \hat{A} \rangle  \quad\quad\quad\quad\quad \quad \frac{d \hat{\rho}}{dt}=-\frac{i}{\hbar}[\lambda q \hat{A},\hat{\rho}],
\end{equation} where we omit the trivial evolution in $q$. Comparing to \eqref{eq: semi_classical_einstein}, we thus see that semi-classical dynamics derived by assuming factorisation are an exact analogue of those commonly used in semi-classical gravity, as has been noted previously \cite{boucher1988semiclassical,HalliwelDH}. This should be expected, given that typical derivations of mean-field semi-classical gravity, such as via the Born-Oppenheimer or effective action approaches, amount to this same approximation \cite{Kiefer:2004xyv,mukhanov2007introduction}. 

The problem is that assuming  factorisation for all times is a \textit{wild} assumption. Namely, this will only hold when the $Q$ subsystem starts in an eigenstate $|a_0\rangle$ or $|a_1\rangle$ of the operator $\hat{A}$. In all other cases, the unitary dynamics of \eqref{eq: toy_model_unitary_dynamics} generates entanglement between the two systems
\begin{equation} \label{eq: entanglement}
    e^{-\frac{i}{\hbar}{\lambda \hat{A}\hat{q}t}}(|a_0\rangle + |a_1\rangle)\otimes |p\rangle = |a_0\rangle \otimes |p- a_0 \lambda t\rangle + |a_1\rangle \otimes |p- a_1 \lambda t\rangle
\end{equation} which violates the factorisation assumption. This means that the standard mean-field semi-classical approach is not a valid approximation when quantum effects on $Q$ are significant, which is precisely the regime semi-classical physics aims to study. 

\subsection*{Failure of semi-classicality in closed systems}

Rather than make the factorisation approximation, one may instead ask why the dynamics \eqref{eq: toy_model_unitary_dynamics_partial_Wigner} itself cannot be used as a semi-classical description.

The answer comes from studying the positivity of $\hat{\varrho}^W(q,p)$. When $W(q,p)\geq 0$, the Wigner function takes the form of a classical probability distribution. Any evolution law that preserves this positivity in time can then be unravelled in terms of individual classical trajectories. One such example is provided by \eqref{eq: semi_classical_liouville}, which is positive and thus can be unravelled as in \eqref{eq: toy_model_standard_semi-classical}. In the same way, $\hat{\varrho}^W(q,p)\succeq 0$ defines a classical-quantum state, the hybrid version of the classical probability and quantum density matrix. Any evolution law that preserves the positivity of this object in time can also be unravelled, this time in terms of individual classical and quantum trajectories. However, it is simple to show that the dynamics of \eqref{eq: toy_model_unitary_dynamics_partial_Wigner} do \textit{not} preserve the positivity of $\hat{\varrho}^W(q,p)$ \cite{boucher1988semiclassical,layton2024classical}. 

The ultimate reason for this is that the original quantum dynamics are highly non-classical on both systems. Indeed, as illustrated in \eqref{eq: entanglement}, the two systems rapidly become maximally entangled for states localised in momentum. Since a maximally entangled state can violate a Bell test, its statistics necessarily cannot be replicated by a classical-quantum state, which in turn implies that $\hat{\varrho}^W(q,p)$ cannot be positive semi-definite.


We thus see that any semi-classical approach starting with \eqref{eq: toy_model_unitary_dynamics_partial_Wigner} was doomed to fail. Since the dynamics generate large amounts of entanglement and are not positive, the system cannot be described in terms of interacting classical and quantum systems. This means the picture \eqref{eq: QQ->CQ} breaks down. The factorisation assumption saves the semi-classical description in this setting, but at the cost of reducing the regime of validity to trivial cases.




\subsection*{An open system semi-classical approach}

Given that the semi-classical formalism breaks down in closed systems, it is reasonable to ask  whether coupling to an environment could limit entanglement generation and hence lead to an emergent classical-quantum description. Such an open system approach is especially natural given that decoherence is understood as the main mechanism for the quantum-to-classical transition \cite{zurek1991decoherence,schlosshauer2007decoherence}.

Let us reconsider the same toy model, but now in an open quantum system setting. Concretely, we assume the bipartite quantum system evolves under
\begin{equation} \label{eq: toy_model_open_dynamics}
    \frac{d\hat{\rho}_{CQ}}{dt}=-\frac{i}{\hbar}[\lambda \hat{q} \hat{A},\hat{\rho}_{CQ}] - \frac{\gamma_C}{\hbar^2}[\hat{q},[\hat{q},\hat{\rho}_{CQ}]] -\gamma_Q[\hat{A},[\hat{A},\hat{\rho}_{CQ}]]. 
\end{equation}  Here $\gamma_C$ and $\gamma_Q$ are positive numbers characterising the decoherence rates on the $C$ and $Q$ subsystems. Such dynamics will generically arise by tracing out a high temperature environment, which in the Markovian limit leads to Lindblad dynamics as given above \cite{breuer2002theory}. In the partial Wigner representation, this can be shown \cite{layton2024classical} to take the form
\begin{equation} \label{eq: toy_model_open_dynamics_partial_Wigner}
    \frac{\partial \hat{\varrho}^W}{\partial t}=-\frac{i}{\hbar}[\lambda q \hat{A}, \hat{\varrho}^W] + \frac{\lambda}{2}\{\hat{A},\frac{\partial\hat{\varrho}^W}{\partial p}\}_+ +\gamma_C\frac{\partial^2 \hat{\varrho}^W}{\partial p^2} -\gamma_Q[\hat{A},[\hat{A},\hat{\varrho}^W]] ,
\end{equation} which provides an entirely equivalent description of the open quantum dynamics \eqref{eq: toy_model_open_dynamics} in a partial phase space picture.

\begin{figure}[]
    \centering
    \includegraphics[width=0.8\textwidth]{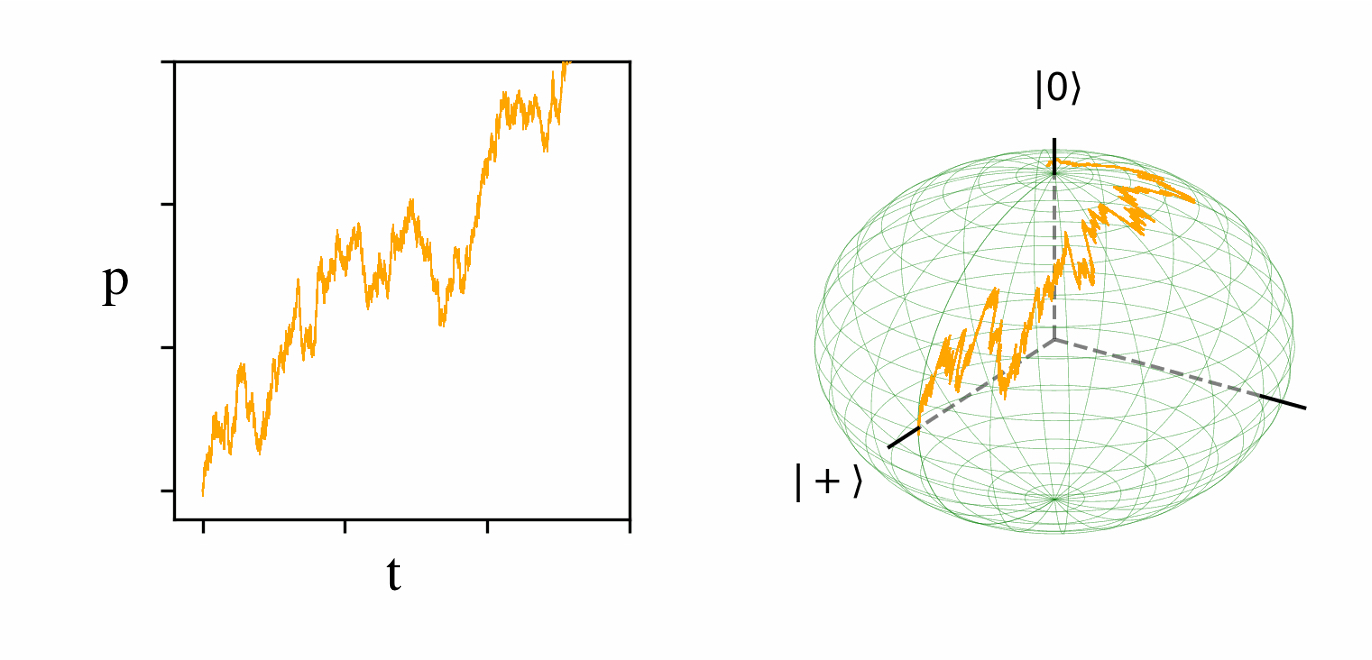}
    \caption{An illustration of a typical trajectory found by solving \eqref{eq: toy_model_improved_semi_classical} with $\hat{A}=\hat{\sigma}_z$ and $\lambda=-1$. On the left, the effective classical particle's momentum is plotted as a function of time, while the right gives the trajectory of the quantum state on the Bloch sphere. For 50\% of noise realisations, an initial superposition of states $|+\rangle=\frac{1}{\sqrt{2}}(|0\rangle + |1\rangle)$ collapses to the $|0\rangle$ state, with the effectively classical particle being pushed upwards as shown. The other 50\% show a downwards drift and collapse to the $|1\rangle$ state.}
    \label{fig: trajectory}
\end{figure}

Rather than make an approximation, as in the standard semi-classical approach, we now employ a tool from the theory of hybrid classical-quantum master equations \cite{diosi1995quantum,diosi2014hybrid,oppenheim2023postquantum,DiosiHalliwel,layton2022healthier}. Namely, we note that when the diffusion, back-reaction and decoherence coefficients obey the relation 
\begin{equation} \label{eq: dec_diff_trade_off}
    \gamma_C \gamma_Q \geq \frac{\lambda^2}{16}
\end{equation} then \eqref{eq: toy_model_open_dynamics_partial_Wigner} preserves the positivity of $\hat{\varrho}^W(q,p)$ and may be unravelled as
\begin{equation} \label{eq: toy_model_improved_semi_classical}
    \frac{dp}{dt}=-\lambda \langle \hat{A} \rangle + \sqrt{\gamma_C}\xi_t \quad \quad \quad \frac{d \hat{\rho}}{dt}=-\frac{i}{\hbar}[\lambda q \hat{A},\hat{\rho}] -\gamma_Q[\hat{A},[\hat{A},\hat{\rho}]] +\frac{\lambda}{\sqrt{8\gamma_C}} \big(\hat{A}\hat{\rho}+\hat{\rho}\hat{A}-2\langle \hat{A} \rangle \hat{\rho}\big)\xi_t
\end{equation} where $\xi_t$ is a white noise process satisfying $\mathbb{E}[\xi_t \xi_s]=\delta(t-s)$. An illustration of a typical trajectory of \eqref{eq: toy_model_improved_semi_classical} is plotted in Fig \ref{fig: trajectory}. Averaging over many trajectories recovers the original bipartite quantum dynamics in the partial Wigner representation via the expression ${\hat{\varrho}^W(q,p)=\mathbb{E}[\hat{\rho}_t \delta(p_t-p)\delta(q_0-q)]}$, where $\rho_t,p_t$ are the solutions to \eqref{eq: toy_model_improved_semi_classical} and $q_0$ is a random variable corresponding to the initial position.

There are three main takeaways from this semi-classical theory. First, it is only valid when \eqref{eq: dec_diff_trade_off} holds i.e. when the product of environmental decoherence rates of the subsystems is sufficiently large compared to their coupling squared. Second, it provides a modified, stochastic version of the standard mean-field approach: comparing \eqref{eq: toy_model_standard_semi-classical} and \eqref{eq: toy_model_improved_semi_classical}, we see the open system approach takes the same form as the standard one, with additional terms characterised by the environmental decoherence rates $\gamma_C$ and $\gamma_Q$. Finally, in contrast to the standard semi-classical approach, this semi-classical dynamics is an exact description of the original quantum dynamics \eqref{eq: toy_model_open_dynamics} \textit{without any approximation}.

\subsection*{Regime of validity of the standard semi-classical approach}

Using this exact semi-classical approach, we can study in which regimes the standard mean-field approach can be used to approximate the open quantum system dynamics.

To see this, we appeal to the argument given in \cite{layton2022healthier}. If we wish for the classical evolution in \eqref{eq: toy_model_improved_semi_classical} to appear deterministic as in \eqref{eq: toy_model_standard_semi-classical}, we require the noise to be insignificant. This requires observing the system over timescales
\begin{equation}
    \tau \gg \frac{\gamma_C}{\lambda^2 \langle \hat{A} \rangle^2},
\end{equation} such that we do not notice the noise in comparison to the much larger force on the system. On the other hand, to recover the quantum dynamics in \eqref{eq: toy_model_standard_semi-classical} from that in \eqref{eq: toy_model_improved_semi_classical}, we must observe the system over timescales short compared to the decoherence time i.e.
\begin{equation}
    \tau \ll \frac{\gamma_Q^{-1} }{\langle\hat{A}^2\rangle- \langle \hat{A} \rangle^2}.
\end{equation} Combining these bounds gives 
\begin{equation} \label{eq: standard_semi_classical_inequalities}
    \langle \hat{A} \rangle^2 \gg \frac{ \gamma_C}{\lambda^2 \tau}\geq \frac{1}{16 \gamma_Q \tau} \gg \langle\hat{A}^2\rangle- \langle \hat{A} \rangle^2,
\end{equation} where the middle inequality uses \eqref{eq: dec_diff_trade_off}. 

This set of inequalities fully characterise the timescales and states for which \eqref{eq: toy_model_standard_semi-classical} approximates \eqref{eq: toy_model_open_dynamics}. For any timescale $\tau$ to exist, the $Q$ system's state must therefore satisfy $\langle \hat{A} \rangle^2 \gg \langle\hat{A}^2\rangle- \langle \hat{A} \rangle^2$
i.e. the quantum state must be highly peaked in the coupling operator $\hat{A}$. This is the analogue of the well-known postulated condition for using \eqref{eq: semi_classical_einstein} to study semi-classical gravity \cite{Ford:1982wu,Kuo:1993if}. However, \eqref{eq: standard_semi_classical_inequalities} illustrates that constraining states alone is not sufficient: one must also restrict the timescale over which the system is observed. For a given state, \eqref{eq: standard_semi_classical_inequalities} shows that the available timescales depend on the decoherence rates on the $C$ and $Q$ degrees of freedom.

By contrast, the full semi-classical theory we provide in \eqref{eq: toy_model_improved_semi_classical} is an exact description of the quantum system \eqref{eq: toy_model_open_dynamics} for all states and timescales. Our approach thus naturally extends the regime of validity of the standard mean-field one.

\subsection*{Discussion}

We have argued that without including open system effects, semi-classical approaches will necessarily fail to accurately approximate the original quantum dynamics in regimes of interest. By contrast, approaches that correctly include decoherence from an environment can lead to semi-classical models that are exact descriptions of the quantum dynamics, without any approximation. 

A simple example of where including open system effects is both natural and necessary is provided by the infamous Page-Geilker experiment \cite{page1981indirect}. Here the positions of two macroscopic lead masses were set depending on an atomic radioactive decay process, and the corresponding gravitational field was measured. Naively applying a semi-classical theory of the kind of \eqref{eq: toy_model_standard_semi-classical} to the extended quantum system of the macroscopic masses, the Geiger counter system and their environment implies that the gravitational field should be sourced by the average configuration of the macroscopic masses. However, a more reasonable approach would model only the macroscopic masses, and treat their coupling to the environment via a large decoherence rate $\gamma_Q$. Applying the semi-classical theory of \eqref{eq: toy_model_improved_semi_classical}, with $\gamma_C$ taken to be a small number such that \eqref{eq: dec_diff_trade_off} is satisfied, one predicts a classical mixture of gravitational fields, as actually observed in the experiment. 

The toy model discussed here misses some of the complexities of the general theory in \cite{layton2024classical}. For quantum models that are not linear, an $\hbar \rightarrow 0$ limit is necessary in order to ensure a consistent semi-classical theory. While here we assume independent decoherence channels on the $C$ and $Q$ systems, these may also be obtained by considering sufficiently strong decoherence on $C$ alone. Although here the partial Wigner function is used for simplicity, a stricter notion of classicality is based on the partial $P$ function, the positivity of which guarantees that the entanglement between the $C$ and $Q$ systems is zero. Putting these details to one side, the message remains that bipartite quantum systems give effective classical-quantum models, and hence consistent semi-classical theories, under sufficiently strong environmental decoherence.

Of course, there is no guarantee that a given bipartite quantum system is truly decohered in the way postulated in \eqref{eq: toy_model_open_dynamics} with the additional condition \eqref{eq: dec_diff_trade_off}. In the case of semi-classical gravity, it may be that while $\gamma_Q$ corresponding to the decoherence of matter is significant, $\gamma_C$ itself is sufficiently small that \eqref{eq: dec_diff_trade_off} does not hold. Indeed, recent proposals for ruling out the possibility of classical gravity amount to checking whether $\gamma_C$ indeed satisfies this relation \cite{oppenheim2023postquantum,oppenheim2023gravitationally,hirotani2026testing,fabiano2026minimal}. 

In the case that \eqref{eq: dec_diff_trade_off} does not hold, the semi-classical interpretation of \eqref{eq: toy_model_improved_semi_classical} breaks down, with the quantum state $\hat{\rho}$ no longer guaranteed to be positive semi-definite. In these regimes, the quantum dynamics cannot be treated as effectively classical-quantum. However, theories in which one system is described in phase space may still be useful, provided an $\hbar\rightarrow 0$ limit is justified. This suggests that semi-classical evolution laws, such as \eqref{eq: toy_model_improved_semi_classical}, may still provide a useful approximation of the original quantum dynamics, even when environmental decoherence is too weak to guarantee true emergent classicality in a subsystem. 

\subsubsection*{Acknowledgements} IL thanks Jonathan Oppenheim and the attendees of “Concepts of Quantum and Spacetime (KEK, Japan, March 2026)” for helpful discussions relating to this work.

\bibliography{bibliography}
\bibliographystyle{unsrt}

\end{document}